\documentclass[aps,twocolumn,floats,prb]{revtex4}
\usepackage{graphics,dcolumn}

\begin{document}

%======================================================================
\title{A unified electrostatic and cavitation model for first-principles
molecular dynamics in solution}
%======================================================================

\author{Dami\'{a}n A. Scherlis$^\ddag$}
\author{Jean-Luc Fattebert$^\dag$}
\author{Fran\c{c}ois Gygi$^\dag$}
\author{Matteo Cococcioni$^\ddag$}
\author{Nicola Marzari$^{\ddag}$}
\affiliation{$^\ddag$Department of Materials Science and Engineering, 
Massachusetts Institute of Technology, Cambridge MA 02139,}
\affiliation{$^\dag$Center for Applied Scientific Computing, Lawrence
Livermore National Laboratory, Livermore CA 94551}

\begin{abstract}

The electrostatic continuum solvent model developed by Fattebert and
Gygi is combined with a first-principles
formulation of the cavitation energy based on a natural
quantum-mechanical definition for the
surface of a solute. Despite its simplicity, the cavitation
contribution calculated by this approach is found to be
in remarkable agreement with that obtained by more complex
algorithms relying on a large set of parameters.
Our model allows for very efficient Car-Parrinello simulations
of finite or extended systems in solution, and demonstrates a
level of accuracy as good as that of established
quantum-chemistry continuum solvent methods.
We apply this approach to the study
of tetracyanoethylene dimers in dichloromethane, providing
valuable structural and dynamical insights on the
dimerization phenomenon.
\end{abstract}

\date{\today}
\maketitle

\section{Introduction}

The importance of electronic structure calculations in solution
is self-evident: chemistry in
nature and in the laboratory often takes place in water or
other solvents, or at a solid-solvent interface.
This is true for all of biochemistry, for most of organic, inorganic, and 
analytical chemistry, and for a vast part of materials and
surface sciences. The natural solution to this problem is
to explicitly include the solvent molecules in the system,
either as one or several solvation shells or as a bulk medium that fills
the simulation
box in periodic boundary conditions. Such approach rapidly increases
the expense of the calculation and is not always affordable. The reasons
are twofold: the cost of an electronic-structure calculation scales as 
the cube of the number of atoms considered, at fixed density. Also,
one needs to ensure that the solvent is treated
appropriately as a liquid medium, using e.g. extensive Monte Carlo
or molecular dynamics simulations. Given the large
ratio between the number of degrees of freedom in the solvent vs. the solute,
the statistical accuracy needed makes most of these approaches prohibitively
expensive. The use of hybrid quantum-mechanical/molecular-mechanics 
(QM/MM) techniques,\cite{merz}$^-$\cite{jpcqmmm} 
in which the solvent atoms are represented with point 
(or Gaussian) charges and classical potentials,
can sensibly alleviate the cost of the computations,
but does not remove the requirement of
long dynamical trajectories of the combined quantum and classical fragments
to simulate the liquid state of the
solvent and to extract thermodynamical averages.

Alternative to these explicit approaches, a description of the solvent 
as a continuum dielectric medium surrounding a quantum-mechanical 
solute has long been established, and has
proved efficient and accurate in a diversity 
of cases.\cite{levine}$^-$\cite{crtruhlar} 
In continuum schemes the dielectric fills
the space outside a cavity where the solute is confined; the shape
of this cavity, considered as a single sphere\cite{onsager}
or ellipsoid in early implementations, has evolved to 
more realistic molecular shapes such as those
defined by interlocking spheres centered on the atoms or by 
isosurfaces of the electron density.\cite{levine,crtomasi1}
In the context of continuum models
the interaction between the dielectric medium and the charge distribution
of the solute provides the electrostatic part of the solvation
free energy, $\Delta G_{el}$, which is the dominant contribution for
polar and charged solutes. 
Solvation effects beyond electrostatic screening, conventionally 
partitioned in cavitation, dispersion, and repulsion,\cite{crtomasi1}
are also important and will be discussed in the context
of our model in Section II. In principle, the application of continuum
models demands that no strong specific
interactions are present between the solvent and the solute molecules,
although the solvent can always be reintroduced explicitly as an
``environmental'' skin for the first solvation shells.

Inexpensiveness is not the single asset of continuum models against
explicit solvent methods. Unless Monte Carlo or molecular
dynamics techniques are used, 
it is unclear what orientation to choose for the
molecules in an explicit solvent model, and even for a medium-sized
solute there may be a large number of possible configurations 
with multiple local minima.\cite{crtruhlar}
More importantly, geometry relaxations will describe a solid or 
glassy phases for the solvent, with a mostly electronic
dielectric screening 
that may differ substantially from its static limit. This 
is particularly true for water, where the
static permittivity $\epsilon_0$ of the liquid is 
larger by a factor of twenty than its electronic $\epsilon_\infty$
contribution. When geometry optimizations including many solvent
molecules are performed, changes
in the solute---e.g. the hydration energy---remain ``buried'' or hidden 
by the large contributions coming from the energy of the solvent.
To extract meaningful information,
Monte Carlo or molecular dynamics simulations with accurate 
thermalizations and averaging times are necessary.
Still, it is far from clear that even first-principles molecular
dynamics treatments of a solvent would provide the accuracy needed
to reproduce static screening as a function of temperature (as an example,
the dielectric constant of water varies between 87.8 at 0 $^{\circ}C$ 
and 55.8 at 100 $^{\circ}C$). Room temperature is well below
the Debye temperature of many solvents, and thus the effect of 
quantum, Bose-Einstein statistics can be very important. In fact,
recent first-principles molecular dynamics studies
of water point to the fact that a combination of
inaccuracies in the quantum-mechanical models (such as density-functional
theory in generalized-gradient approximations) and the use of Boltzmann
statistics produce an overstructured description of water
\cite{pattrick}$^-$\cite{galli2}, with apparent freezing roughly
a hundred degrees above the experimental point.
Last, the relaxation times needed to extract thermodynamical data
from a solvated system can be exceedingly long,\cite{pasquarello}
compounding many of the issues highlighted here
(dynamical, as opposed to static screening, would require to take into account
the solvent relaxation times, either explicitly or via a frequency-dependent
dielectric model, but such a framework goes beyond the scope of this paper).
Continuum solvent methods are free from these issues, and for this
reason alone they may be the first
choice even when computational resources are not the main constraint.

The presence of a polarizable 
dielectric will induce a charge redistribution in the solute, which
in turn will affect the polarization of the medium. In
the self-consistent reaction field approach (SCRF) 
the dielectric medium and the electronic density
respond to the electrostatic field of each other in a self-consistent 
fashion.\cite{levine} Over the past twenty five years
a number of developments stemming from the SCRF approach
have been proposed and further elaborated.
\cite{tomasi1}$^-$\cite{jensen}
Among these, the Polarizable-Continuum Model (PCM) of Tomasi et al.
\cite{crtomasi, tomasi1, cpl-cossi} and the Conductorlike Solvation Model
(COSMO) of Klamt and Sch\"{u}\"{u}rmann\cite{cosmo}
are probably the most-widely used choices in quantum chemistry applications.
In both cases the dielectric constant $\epsilon$ is
taken to be 1 inside the cavity, and a fixed value outside
(equal to the dielectric constant of the solvent for PCM,
or infinite for the case of COSMO).
The electrostatic problem is then formulated in terms of
apparent surface charges (ASC) distributed on the solute-solvent interface.
For first-principles molecular dynamics applications,
the discontinuity of $\epsilon$ at the interface 
needs to be removed to calculate accurately 
the analytic derivatives of the potential with respect to the
ionic positions. This may be accomplished with
the use of a smoothly varying dielectric potential that restores
well-behaved analytic gradients.\cite{karplus} Still,
Born-Oppenheimer ab-initio molecular dynamics in localized basis
sets are demanding enough
that they have yet to be combined, to the best of our knowledge,
with the ASC approach for realistic simulations of medium or large system.

On the other hand,
first-principles implementations of the continuum solvent model within
the Car-Parrinello framework\cite{carparrinello} have been
devised,\cite{cpldangelis}$^-$\cite{arias} even though dynamical
studies have been reported, to the best of our knowledge,
in only few cases.\cite{jluc2,ziegler}
In this paper, we 
introduce a first-principles and conceptually simple approach to the
calculation of cavitation energies based on the 
definition of a quantum surface for the solvent.\cite{matteo}
We combine this scheme with the electrostatic 
solvation model of Fattebert and Gygi,\cite{jluc2,jluc1}, and find
a level of accuracy at least as good as that of established
quantum-chemistry treatments. 
The model requires no adjustable parameters
other than a universal definition of the
cavity (practically depending on one parameter),
and the dielectric constant and the surface tension of the solvent.
This combined model is well
suited for first-principles molecular dynamics calculations of 
large finite and extended systems, using e.g.
efficient plane-wave Car-Parrinello implementations.
In the following
sections we describe the method and  examine its performance in comparison
with experiments and with the well-established PCM approach. Finally, given
that cavitation contributions can be
particularly important in dimerization processes
(where the fusion of two cavities into one provides an additional
stabilizing energy), we employ our method
to study the association of the tetracyanoethylene
(TCNE) anion in solution\cite{kochi} by means of static and dynamical 
simulations, highlighting the role of the cavitation term in the dimerization.

\section{The model and its context}

\subsection{Preliminary details}

Our continuum solvation model has been implemented in the
public domain Car-Parrinello parallel code included in the Quantum-ESPRESSO
package,\cite{Espresso} based on density-functional theory (DFT),
periodic-boundary conditions, plane-wave basis sets, and
pseudopotentials to represent the ion-electron interactions.
All calculations reported in this work, unless otherwise noted,
have been performed using
Vanderbilt ultrasoft pseudopotentials,\cite{usp} with the
Kohn-Sham orbitals and charge density expanded in plane 
waves up to a kinetic energy cutoff of 25 and 200 Ry respectively.
In the Appendix we review the formalism used
to calculate energies and forces in periodic boundary conditions
in the context of our implementation.
Further details can be found in reference~~\cite{pasq-galli}.

We adopt the definition introduced by Ben-Naim for the solvation free energy,
\cite{bennaim}
in which $\Delta G_{sol}$ corresponds to the process of transferring
the solute molecule from a fixed position in the gas phase to a fixed
position in the solution at constant temperature, pressure, and
chemical composition. For calculation purposes and especially in the
case of the continuum dielectric model,
$\Delta G_{sol}$ can be regarded as the sum of several
components, of which the electrostatic, the cavitation, and
the dispersion-repulsion contributions are the most relevant
($\Delta G_{sol} = \Delta G_{el}+\Delta G_{cav}+
\Delta G_{dis-rep}$).\cite{nota1}
None of these, however, can be directly obtained through
experiment, the sum of all of them, $\Delta G_{sol}$, being
the only measurable quantity. In our model,
$\Delta G_{el}$ and $\Delta G_{cav}$ are considered explicitly,
while $\Delta G_{dis-rep}$, less relevant for the systems
considered here, is largely seized by virtue of the
parametrization, as part of the electrostatic term. 
The dispersion-repulsion energy
may be important in the case of hydrophobic
and aromatic species, but its explicit calculation is beyond
the aim of the present work---in particular, the implementation
of the technique proposed by Floris, Tomasi and
Pascual Ahuir\cite{disp1,disp2} would be straightforward in
our model.

\subsection{Electrostatic solvation energy}

The electrostatic interaction between the dielectric and the solute
is calculated as proposed by Fattebert and Gygi.\cite{jluc2,jluc1}
In the following we provide an outline of the model.

The Kohn-Sham energy functional\cite{parryang} of a system of ions and 
electrons can be written as
\begin{equation}
E[\rho] = T[\rho]+ \int v({\bf r}) \rho({\bf r}) d{\bf r} + E_{xc}
+\frac{1}{2} \int \rho({\bf r}) \phi[\rho] d{\bf r}
\end{equation}
where the terms on the right hand side correspond to 
the kinetic energy of the
electrons, the interaction energy with the ionic potential, the 
exchange-correlation energy, and the electrostatic energy $E_{es}$
respectively. In the standard energy functional, the electrostatic
potential $\phi[\rho]$ is the
solution to the Poisson equation in vacuum,
\begin{equation}
\nabla ^2 \phi = -4\pi \rho~.
\end{equation}
In the presence of a dielectric continuum with a permittivity
$\epsilon[\rho]$, the Poisson equation becomes
\begin{equation}
\nabla \cdot (\epsilon[\rho] \nabla \phi) = -4\pi \rho~.
\end{equation}
By inserting the charge density obtained from Eq.~(3) into the
expression for the electrostatic energy, and integrating by parts, we obtain:
\begin{equation}
E_{es} = \frac{1}{8\pi} \int \epsilon[\rho](\nabla \phi[\rho])^2d{\bf r}.
\end{equation}
While Eq.~(2) can be efficiently solved in reciprocal
space with the use of fast Fourier transforms, for arbitrary
$\epsilon[\rho]$ the Poisson equation (3) must be solved
with an alternative numerical scheme.
In the present case, it is discretized on a real space grid,
and solved iteratively using a multigrid
technique.\cite{jluc2} The functional derivative of $E_{es}$ with respect to
$\rho$ yields $\phi$ and an additional
term $V_{\epsilon}$, originating in the dependence of
the dielectric function on the charge density:
\begin{equation}
\frac{\delta E_{es}}{\delta \rho}({\bf r}) = 
\phi({\bf r}) + V_{\epsilon}({\bf r}),
\end{equation}
\begin{equation}
V_{\epsilon}({\bf r})=-\frac{1}{8\pi} (\nabla \phi({\bf r}))^2 
\frac{\delta \epsilon}{\delta \rho}(\bf r).
\end{equation}
The self-consistent Kohn-Sham potential is constructed summing
$V_{\epsilon}$ and the electrostatic
potential $\phi$, to which contributions from the exchange-correlation, and
the local and non-local terms in the pseudopotentials are also added (see Appendix).
The dielectric medium and the
electronic density then respond self-consistently to each other
through the dependence
of $\epsilon$ on $\rho$ and viceversa.

As already mentioned in the introduction, in Gaussian-basis sets
implementations of the continuum model $\epsilon$ is a binary function
with a discontinuity at the cavity surface. The accurate representation
of such a function would require unrealistic high kinetic energy cutoffs for
the plane wave basis and expensive real space grids. The use of smoothly varying 
dielectric functions instead eases the numerical load and
avoids discontinuities in the forces, essential to proper energy conservation
during molecular dynamics simulations. Also, a smooth decay
of the permittivity in the proximity of the
solute-solvent boundary may even be considered a more physical
representation than a sharp discontinuity.
In our implementation the dielectric medium is defined using
two parameters $\rho_0$ and $\beta$:
\begin{equation}
\epsilon(\rho({\bf r})) = 1+ \frac{\epsilon_\infty -1}{2}\left (
1+\frac{1-(\rho({\bf r})/\rho_0)^{2\beta}}{1+(\rho({\bf r})/\rho_0)^{2\beta}}
\right ).
\end{equation}
This function asymptotically approaches $\epsilon_\infty$ 
(the permittivity of the bulk solvent) in regions of space where the
electron density is low, and 1 in those regions where it is high.
The parameter $\rho_0$ is the density threshold determining the cavity
size, whereas $\beta$ modulates the smoothness of the transition from
$\epsilon_\infty$ to 1.

\subsection{Cavitation energy}

The cavitation energy $\Delta G_{cav}$ is defined as the work involved 
in creating the appropriate cavity inside the solution in the 
absence of solute-solvent interactions.\cite{crtomasi}
Different approaches
have been introduced to compute $\Delta G_{cav}$;
nevertheless it is unclear which one is the most accurate given
the unavailability of experimental values to compare.
Formulations based on the scaled particle theory\cite{spt1,spt2}
have been originally proposed by Pierotti\cite{pierotti} 
and further developed in several
studies.\cite{claverie}$^-$\cite{benzi} Although these
approaches are derived from a rigorous statistical mechanics standpoint,
eventually the use of a set of fitted parameters is needed
to represent an effective radius for the solvent and for the spheres
centered on the solute atoms. For nonspherical cavities, one of the
most used approximations is the
so-called Pierotti-Claverie formula:\cite{crtomasi1,claverie}
\begin{equation}
\Delta G_{cav}= \sum_{k=1}^{N} \frac{A_k}{4\pi R_k^2} G_{cav}(R_k).
\end{equation}
Eq.~(8) describes the cavity as the volume occupied by
$N$ interlocked spheres centered on the atoms;
$A_k$ is the area of atom $k$ exposed to the solvent, $R_k$ is
its van-der-Waals radius, and $G_{cav}(R_k)$ is the cavitation free
energy associated to the creation of a spherical cavity of radius $R_k$
according to Pierotti.\cite{pierotti}

Efforts have also been
made to describe $\Delta G_{cav}$ as a function of the macroscopic
surface tension of the solvent $\gamma$.\cite{uhlig}$^-$\cite{ahuir}
The suggestion of Uhlig\cite{uhlig} of expressing the work 
involved in producing the cavity as the product between $\gamma$
and the area of a sphere, $\Delta G_{cav} = 4\pi R^2 \gamma$,
has been extended to account for the curvature of
the solute-solvent interface, according to the theory of Tolman for
the surface tension of a droplet.\cite{tolman}
The validity of simplified expressions of the kind
\begin{equation}
\Delta G_{cav}=PV + 4\pi R^2 \tilde{\gamma} \left (1-\frac{2\delta}{R} \right )
\end{equation}
has been investigated by different authors\cite{floris,chandler} by means of
Monte Carlo simulations with classical potentials. In Eq.~(9),
$\tilde{\gamma}$ is an effective surface tension for the interface,
$R$ is the radius of the cavity, and $\delta$ is a coefficient that
would correspond to the Tolman length in the case of a macroscopic surface.
Studies from both Floris\cite{floris} and Chandler\cite{chandler} 
groups have shown
that $\tilde{\gamma}$ is essentially indistinguishable from the macroscopic
surface tension of the solvent, $\gamma$. Their simulations
have assigned to $\delta$ a value of 0.0
in TIP4P water,\cite{floris} and of the order
of -0.5$\sigma$ in the case of different Lennard-Jones fluids
($\sigma$ being the Lennard-Jones radius),\cite{chandler}
suggesting that the curvature correction can in practice be
ignored for cavities with radii above only a few Angstroms.

In view of these results, we have chosen to estimate
the cavitation energy as the product between the surface tension and
the area of the cavity,
\begin{equation}
\Delta G_{cav} = \gamma S(\rho_0),
\end{equation}
where $S(\rho_0)$ is the surface of the same cavity employed in the 
electrostatic part of the solvation energy and is defined by an isosurface
of the charge density. As observed by Floris et al.,\cite{floris}
there is always a surface in between
the internal and the solvent accessible surfaces such that the
correction factor $(1-\frac{2\delta}{R})$ reduces to 1,
entailing a linear dependence between $\Delta G_{cav}$ and the
cavity area. We rely on the parametrization of the density threshold
$\rho_0$ to obtain an appropriate surface.

The area of this cavity can be easily and accurately calculated by 
integration in a real-space grid, as the volume of a thin film delimited 
between two charge density isosurfaces, divided by the thickness of this film.
This idea has been originally proposed by Cococcioni et al.\cite{matteo}
to define a ``quantum surface'' in the context of extended electronic-enthalpy
functionals:
\begin{equation}
S(\rho_0)= \int d{\bf r} \left\{ \vartheta_{\rho_0-
\frac{\Delta}{2}} \left[ \rho({\bf r})\right]
-\vartheta_{\rho_0+\frac{\Delta}{2}} \left[ \rho({\bf r})
\right] \right\} \times \frac{|\nabla \rho ({\bf r})|}{\Delta}.
\end{equation}
The finite-differences parameter $\Delta$ determines the separation between two
adjacent isosurfaces, one external and one internal, corresponding to 
density thresholds $\rho_0 - \Delta/2$ and $\rho_0 + \Delta/2$ 
respectively. The spatial distance between these two cavities---or the 
thickness of the film---is given at any point in space by 
the ratio $\Delta/|\nabla \rho|$.
The (smoothed) step function $\vartheta$ is zero in regions of low
electron density and approaches 1 otherwise, and it has been defined 
consistently with the dielectric function of Eq.~(7):
\begin{equation}
\vartheta[\rho({\bf r})]=\frac{1}{2} \left[ 
\frac{(\rho({\bf r})/\rho_0)^{2\beta}-1}
{(\rho({\bf r})/\rho_0)^{2\beta}+1}+1 \right ].
\end{equation}
Note that the volume of the cavity is simply the integral of $\vartheta$ on
all space:
\begin{equation}
V_c(\rho_0) = \int d{\bf r}~\vartheta_{\rho_0}[\rho({\bf r})].
\end{equation}
The functional derivative of $\Delta G_{cav}=\gamma S(\rho)$ 
with respect to the density
gives then the additional contribution to the Kohn-Sham potential,
\[
\frac{\delta \Delta G_{cav}}{\delta \rho}({\bf r})= \frac{\gamma}{\Delta}
\times \left[ \vartheta_{\rho_0-
\frac{\Delta}{2}} \left[ \rho({\bf r})\right]
-\vartheta_{\rho_0+\frac{\Delta}{2}} \left[ \rho({\bf r})
\right] \right] \]
\begin{equation}
\times \left[ \sum_i \sum_j \frac{\partial_i \rho({\bf r})
\partial_j \rho({\bf r}) \partial_i \partial_j \rho({\bf r})}
{|\nabla \rho ({\bf r})|^3} -\sum_i \frac{\partial_i^2 \rho({\bf r})}
{|\nabla \rho ({\bf r})|} \right]
\end{equation}
where the indices $i$ and $j$ run over the $x$, $y$, $z$ coordinates, and
$\partial_i$ indicates a partial derivative with respect to the position.

\begin{figure} \centerline{
\rotatebox{-90}{\resizebox{3in}{!}{\includegraphics{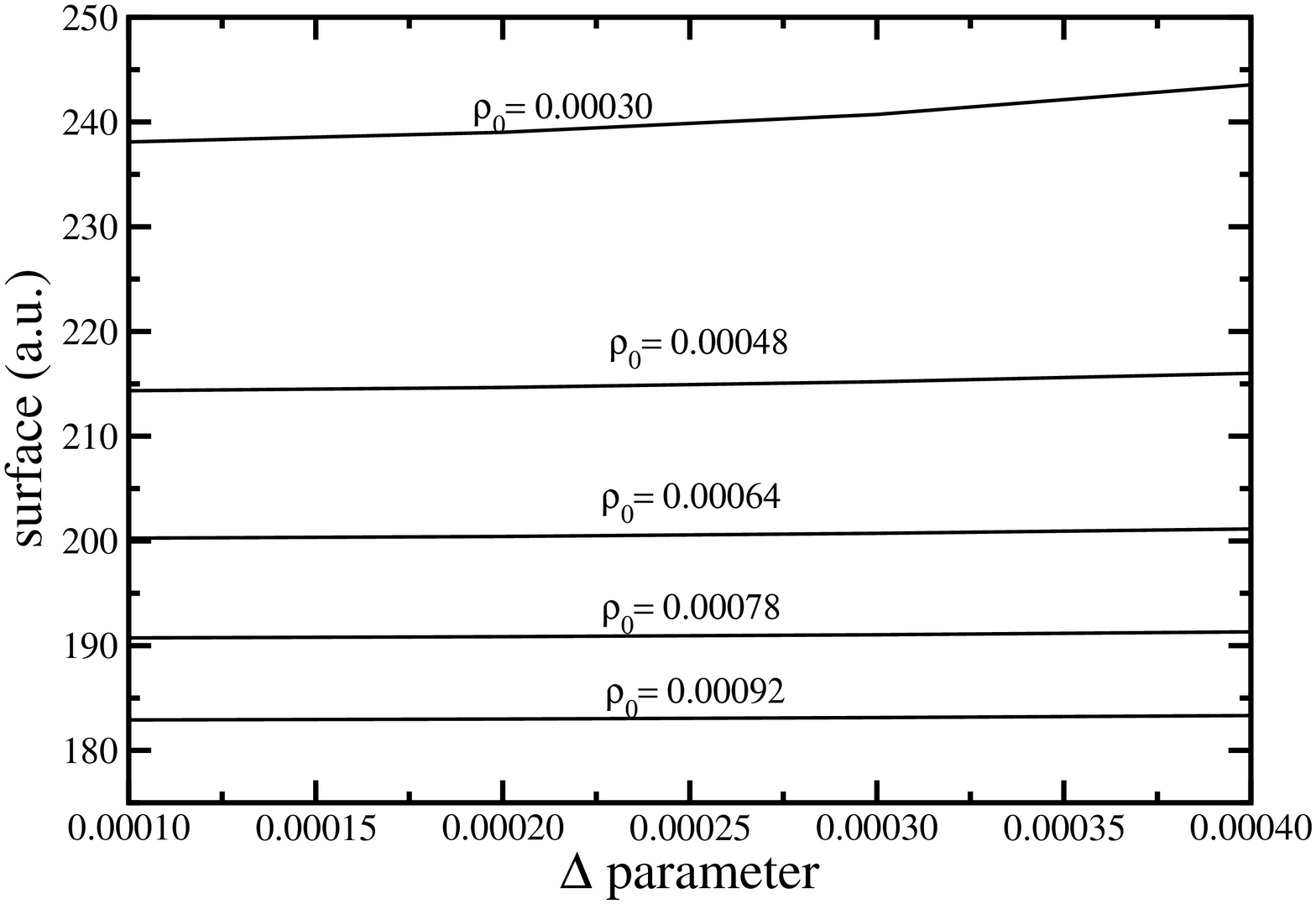}}}
}
\caption{D. Scherlis}
\label{delta}
\end{figure}

The exact value of the discretization $\Delta$ is not important, 
as long as it is chosen
within certain reasonable limits---a very low value would introduce numerical
noise, while a too large one would render an inaccurate measure of the
surface. The freedom in the choice of $\Delta$ is 
illustrated in Fig.~\ref{delta},
where the dependence of $S$ on this parameter is examined for a water molecule
at various thresholds. For $\rho_0$ equal or above 0.00048 e, the calculation
of the cavity area is fairly converged for any value of $\Delta$
within the range displayed in the figure. We have adopted 
a value of $\Delta$=0.0002 e in our simulations. 
It is worth noting, on the other hand, that the dependence of the surface
on the density threshold $\rho_0$ is only moderate, reflecting the
fact that at the ``molecular boundary'',
the electron density decays significantly on a short distance. This behavior
is portrayed in Fig.~\ref{delta}, 
where it can be seen that for a given $\Delta$,
the calculated surfaces change in only about 25\% when $\rho_0$
is increased three times.
$\Delta G_{cav}$ is in fact much less sensitive to the electron density
threshold than $\Delta G_{el}$.

\section{Results and Discussion}

\subsection{Solvation energies in water}

The only adjustable parameters in our solvation model are $\rho_0$ and
$\beta$, which determine the shape of the cavity according to
Eqs. (7) and (12).
Other parameters entering the model, namely the static dielectric constant
and the surface tension of the solvent, are physical constants taken 
from experiments. We have actually kept $\rho_0$ as the single degree of freedom
to fit the solvation energy, while fixing the value of $\beta$ to 1.3 as
in reference~~\cite{jluc1}.
This choice of $\beta$ provides a smooth, numerically convenient transition
for the step function, still ensuring that the lower and upper limits
of $\epsilon(\rho({\bf r}))$ and $\vartheta(\rho({\bf r}))$
are reached reasonably fast.
The parameter $\rho_0$ was obtained from a linear least squares 
fit to the hydration energies of three solutes:
amide, nitrate, and methylammonium (a polar
molecule, and two ions of opposite sign). The resulting value, $\rho_0=$0.00078, 
was employed thereafter in all the simulations. This can be regarded as
a rather universal choice for $\rho_0$ and $\beta$; reparametrizations for
different solvents could be considered (if enough experimental data
were available) probably gaining some marginal
accuracy at the expense of generality.

Table I shows the solvation and cavitation energies in water calculated
for a number of neutral species, along with their experimental
values.\cite{sol1}$^-$\cite{sol3} 
A quite remarkable agreement with experiments is found.
We compare the data with PCM results obtained at the DFT-PBE/6-311G(d,p) level
(or DFT-PBE/3-21G** for the case of Ag$^+$) using 
the Gaussian 03 package.\cite{g03} Also significant
is the accord between the cavitation energies computed with
the two methods---with the caveat that in Gaussian-PCM $\Delta G_{cav}$ 
is based on the Pierotti-Claverie formula (see Eq.~(8)) which 
requires a lengthy list of parameters including all van-der-Waals
radii. Similar agreement between the 
values of $\Delta G_{cav}$ coming our approach and PCM
is found among charged solutes as shown 
in Table II. The level of accuracy in $\Delta G_{sol}$ is in
this case as good as for the neutral
solutes, if viewed in relative terms (we point out that, regarding
the experimental values of $\Delta G_{sol}$ reported for ions,
discrepancies between sources up to a few kcal/mol are common).

\begin{figure} \centerline{
\rotatebox{-90}{\resizebox{3in}{!}{\includegraphics{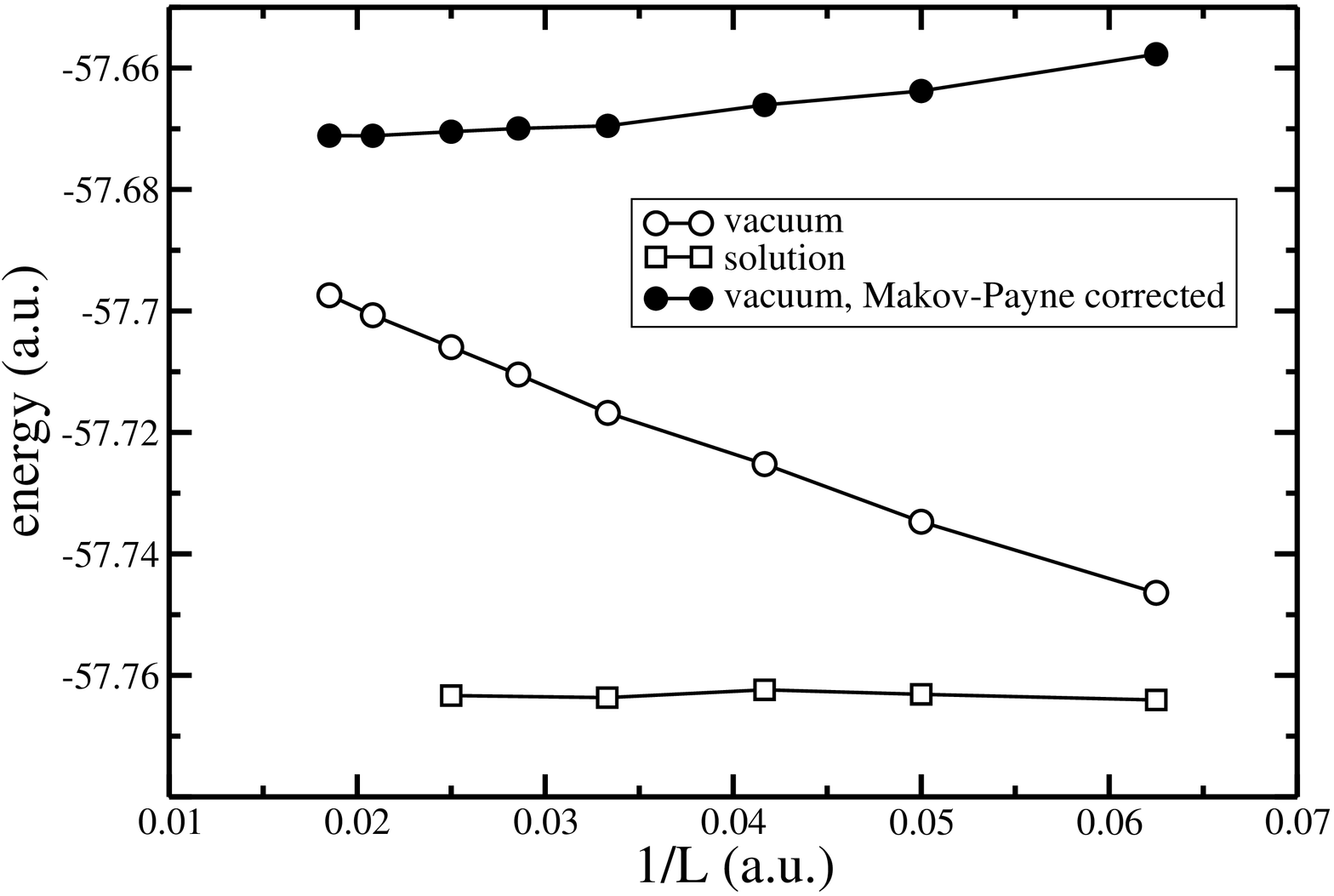}}}
}
\caption{D. Scherlis}
\label{lattice}
\end{figure}

The solvation energies of the ionic solutes showed
in Table II were calculated including the Makov-Payne correction,\cite{makovpayne}
which takes into account how the gas phase energy of a charged system is affected
by its periodic images in supercell calculations:
\begin{equation}
E_{GAS}=E_{PBC}+\frac{q^2 \alpha}{2L}-\frac{2\pi q Q}{3L^3}+O[L^{-5}],
\end{equation}
where $E_{GAS}$ and $E_{PBC}$ are the isolated and the supercell
energies respectively, $q$ is the charge of the system, $Q$ its quadrupole
moment, $L$ the lattice parameter, and $\alpha$ the Madelung
constant (we used a simple cubic lattice of charges, for which
$\alpha$=2.8373\cite{gillan}).
As shown in Fig.~\ref{lattice} for the nitrate anion, the
dependence of the energy with respect to the inverse of the lattice
parameter becomes virtually linear for $L$ above
40 a.u., pointing out that the quadrupole term
can be neglected in supercells of that size or larger. So,
we applied the Makov-Payne correction to the $1/L$ leading order to all the
cations and anions in Table II, always checking for convergence with respect 
to $1/L$. The gas phase energies calculated in this way were 
subtracted from the correspondent energies in solution to obtain
$\Delta G_{sol}$. Fig.~\ref{lattice} also shows
that total energies in solution quickly converge
with respect to the size of the supercell, thanks to
the dielectric screening of the Coulombic interactions
between periodic images.

With the exception of CH$_4$, the solutes in Tables I and II 
are either polar or ionic compounds of relatively small size.  
Since the dispersion-repulsion energy is not explicitly accounted 
for in our model, its accuracy for systems in which this
contribution becomes dominant, such as highly hydrophobic or aromatic 
compounds, will be necessarily affected (this is already the
case for methane).
For the species listed in Tables I and II the dispersion-repulsion effect
is captured to a large extent by our electrostatic term.
As the size of the solute increases and its polarity decreases, though, 
the non-electrostatic
terms tend to monopolize the solvation energy, and a model lacking
the dispersion-repulsion contribution will perform poorly.
This limitation could be possibly overcome by a different parametrization
specific to
large nonpolar solutes, or of course by directly computing the dispersion 
and repulsion contributions.\cite{disp1,disp2}

\subsection{Molecular dynamics and total energy conservation}

\begin{figure} \centerline{
\rotatebox{-90}{\resizebox{3.0in}{!}{\includegraphics{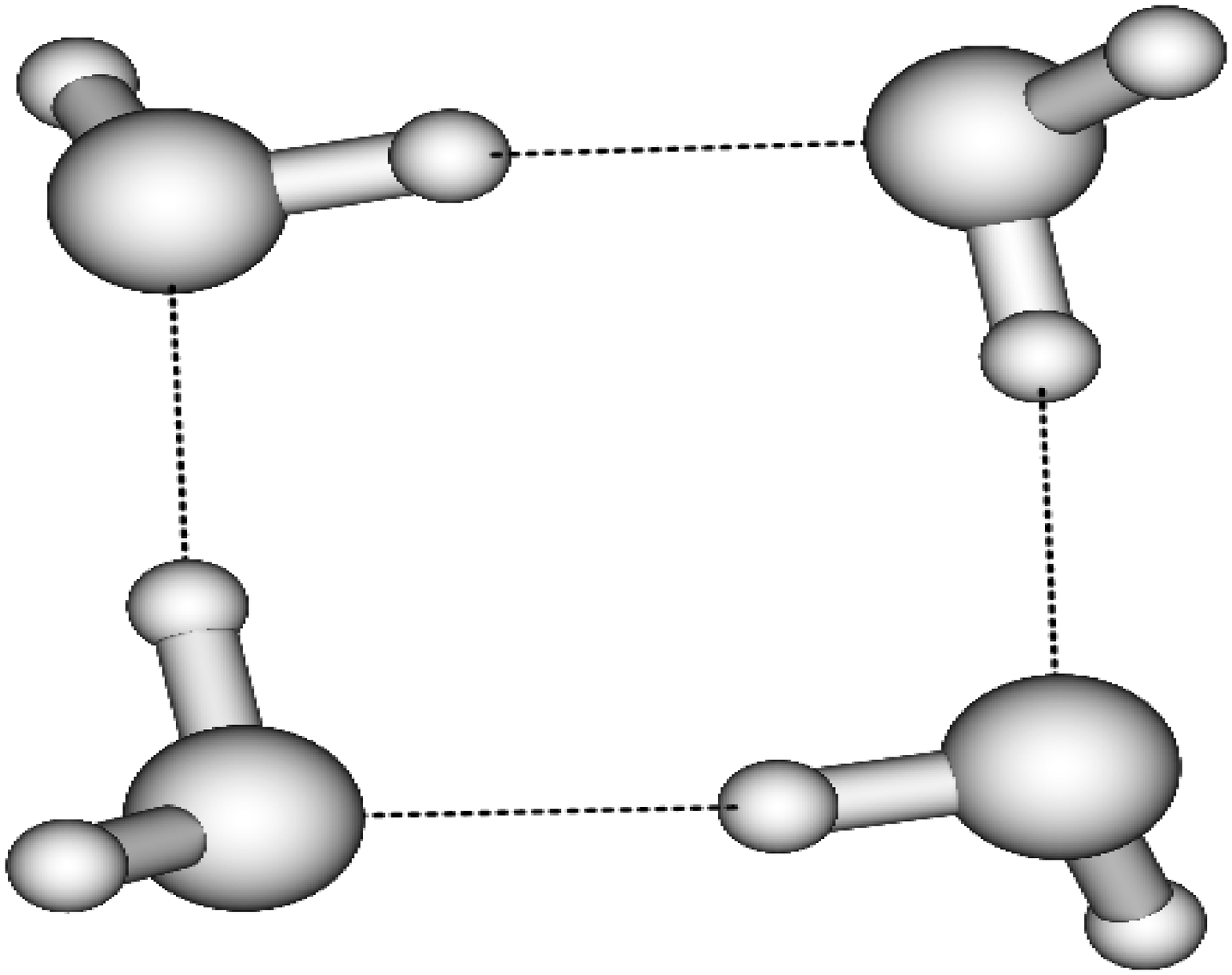}}}
}
\caption{D. Scherlis}
\label{water}
\end{figure}

\begin{figure} \centerline{
\rotatebox{-90}{\resizebox{3in}{!}{\includegraphics{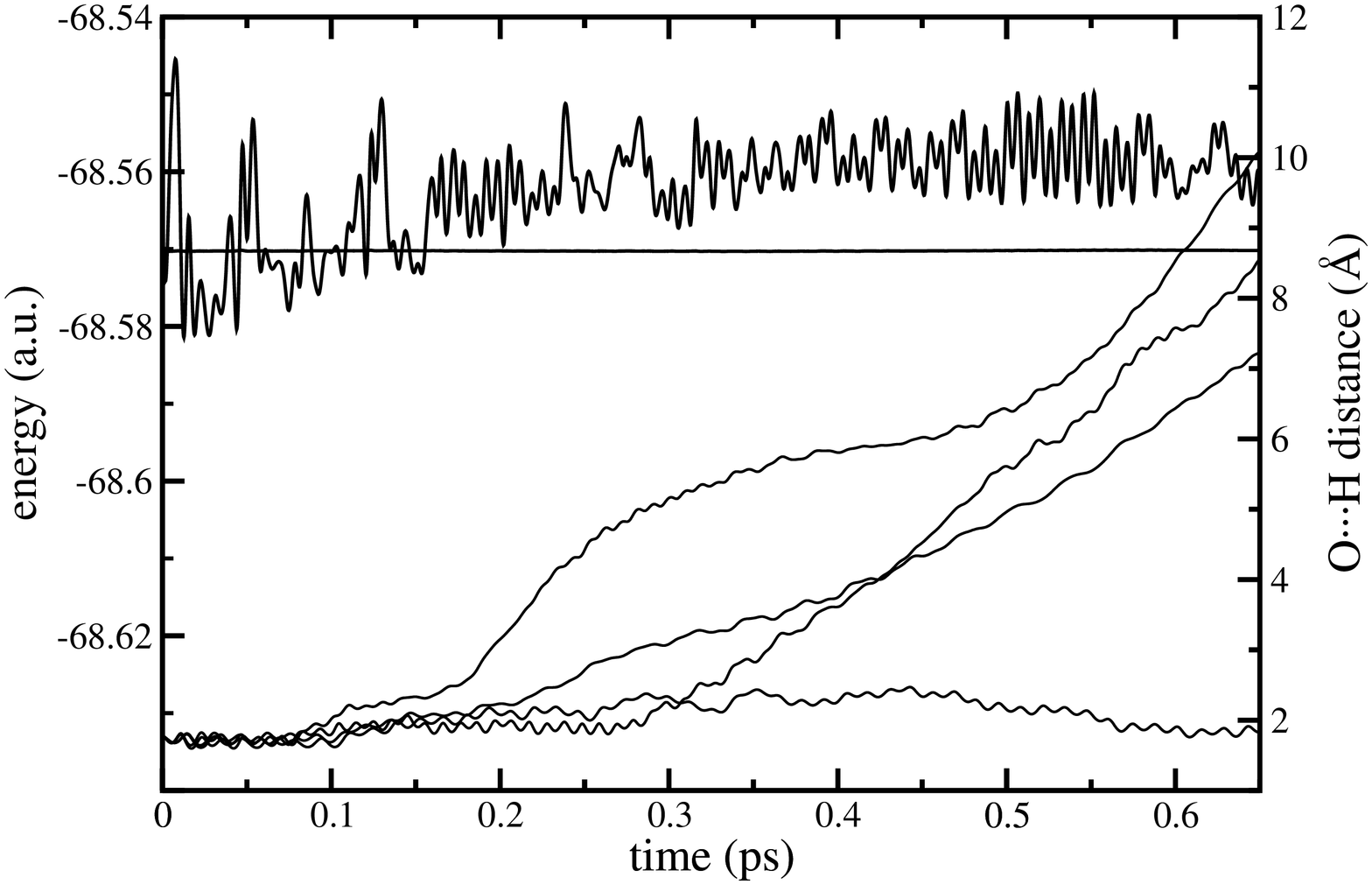}}}
}
\caption{D. Scherlis}
\label{energy}
\end{figure}

As a computational test, we performed
canonical molecular dynamics simulations of a tetramer of D$_2$O molecules
(heavy water), using our solvation-cavitation model to represent an aqueous solution.
A time step of 0.192 fs and an electronic mass of 400 a.u. were employed,
and the system was thermalized at 350 K by applying
the Nose-Hoover thermostat on the ions. Fig.~\ref{water} shows
the initial configuration of the cluster, where the four D$_2$O molecules
are stabilized in a ring by four hydrogen bonds. In the upper part
of Fig.~\ref{energy}, the total energy is monitored
throughout the run and compared with the potential energy.
The conservation of
the total energy is as good as in the gas phase for the same
simulation parameters, and is not affected by the dissociation
of the bonds. The analysis was not pursued beyond 0.65 ps,
when the water cluster dissolves into the medium and one of the D$_2$O molecules
evolves close to the border of the real space grid,
affecting the Dirichlet boundary conditions.

In the lower section of Fig.~\ref{energy}, 
the intermolecular O$\cdots$H distance is plotted for the four initial
hydrogen bonds that keep the cluster bound. The solvent 
dissociates this structure early in the simulation, and before half a
picosecond only one hydrogen bond has survived. By the end of the run
a dimer is what remains of the original tetramer.
For comparison, long molecular dynamics (up to 10 ps)
were carried on in the gas phase under identical conditions. In this case
the cyclic cluster is stable for the full length of the simulation,
showing that the disruption of the intermolecular bonds is indeed a consequence
of the solvation effect.

\subsection{Dimerization of TCNE anions in solution}

Starting in the early 60's, dimerization of charged and
neutral organic $\pi$ radicals in solution and in the solid
state was reported by 
several authors.\cite{spach}$^-$\cite{gundel}
The discovery of this phenomenon prompted a vast amount of
research which has continued up to the present day.
\cite{kochi, sesto1}$^-$\cite{simons} Among the systems addressed,
significant efforts have gone into the study of the tetracyanoethylene
anion [TCNE]$^{-\cdot}$ and 
its salts
because of their central role in the understanding and development
of molecular metals. Recently, evidence has been presented showing that
the dimerization of [TCNE]$^{-\cdot}$ in the solid state involves
two-electron four-center $\pi$*--$\pi$* bonding arising from the interaction
of the two singly occupied molecular orbitals (SOMOs) of the
anions and leading to long ($\approx$ 3.0 $\AA$) intermonomer C--C 
covalent bonds.\cite{sesto1,sesto2} Data from UV-vis and EPR
spectroscopies suggested the same conclusions are true in the
solvated state.\cite{kochi,sesto2} In the gas phase, DFT and MP3 calculations 
show that the dimer is only metastable, since the attractive covalent interaction
between the anions is outweighed by the Coulombic 
repulsion.\cite{sesto1,sesto2,simons} In the solid state, in contrast, the
positive counterions stabilize the array of like charges, allowing the
$\pi$*--$\pi$* bonding to occur.\cite{sesto1,sesto2} 

\begin{figure} \centerline{
\resizebox{3.5in}{!}{\includegraphics{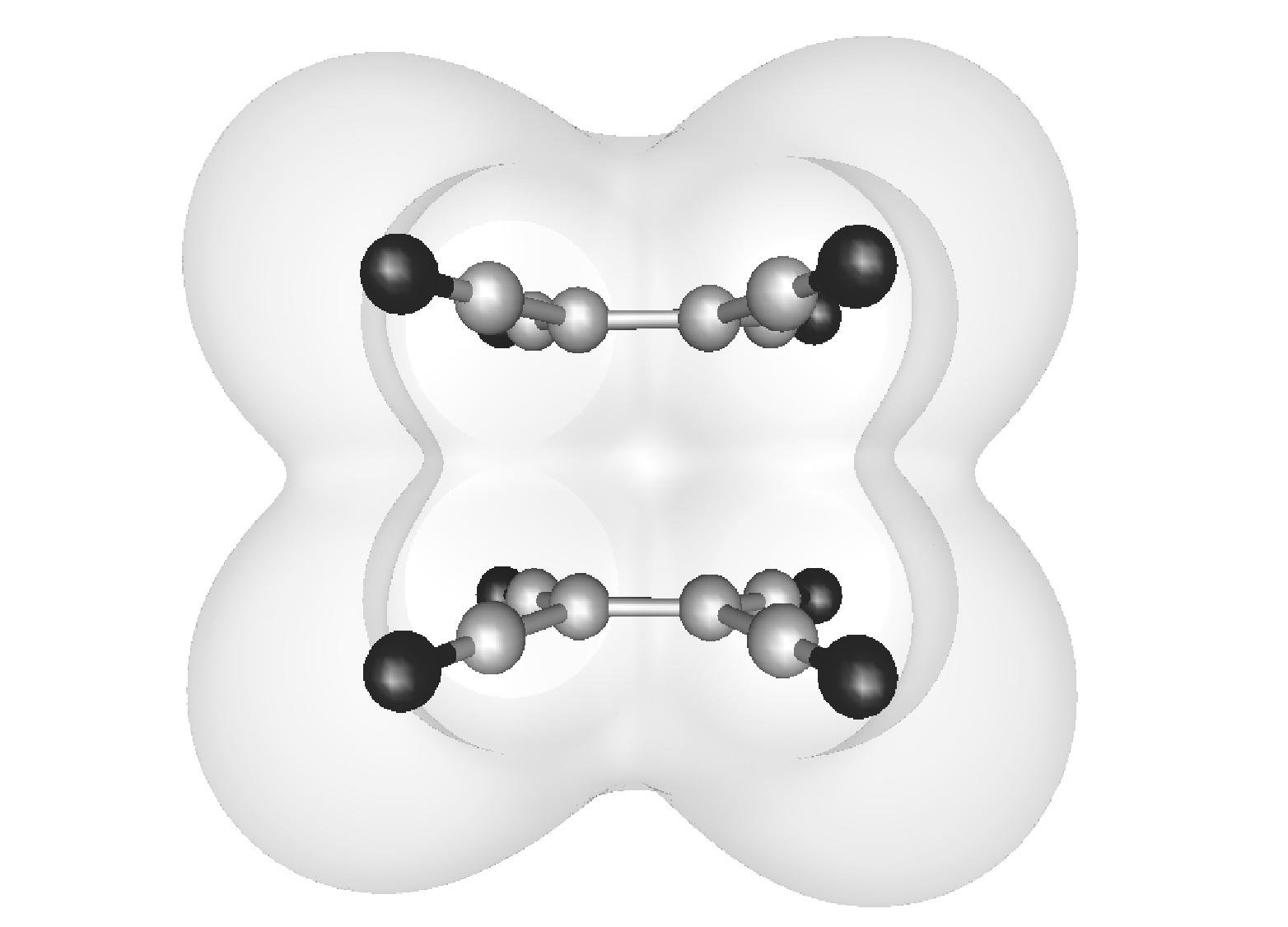}}
}
\caption{D. Scherlis}
\label{tcne}
\end{figure}

A solvent may play
an analogous role in stabilizing the dimer, by favoring the concentration of 
charge in a single cavity. We used our solvation model to fully optimize the 
doubly charged TCNE dimer\cite{nota2} in dichloromethane, properly adapting the
values of $\epsilon$ and $\gamma$ (8.93 and 27.20 mN/m respectively).
We found a stable minimum at 
an equilibrium distance of 3.04 $\AA$, in close agreement with solid state
geometries: X-ray data of different salts\cite{xray1}$^-$\cite{xray3}
range from 2.83 to 3.09 $\AA$. The CN substituents deviate from the
plane by 5$^\circ$ (see Fig.~\ref{tcne}), consistently with
the NC-C-C-CN dihedral angles observed in crystals, between 3.6 and 6.5$^\circ$.
This deviation has been ascribed to the rehybridization of the sp$^2$ carbon
as the intradimer bond is formed,\cite{sesto1} but its origin could be also
tracked to the steric repulsion between the CN moieties facing each other.

\begin{figure} \centerline{
\rotatebox{-90}{\resizebox{3in}{!}{\includegraphics{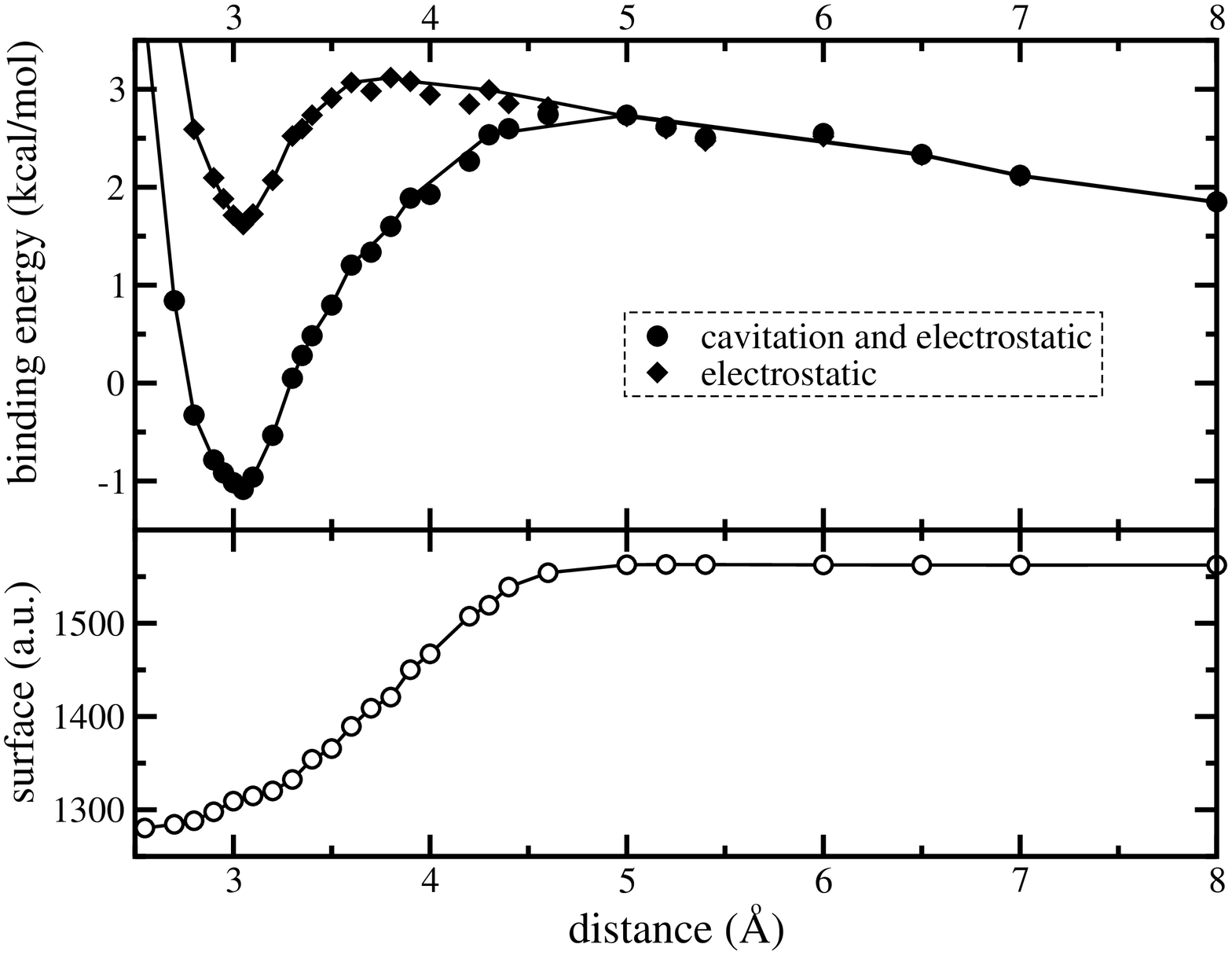}}}
}
\caption{D. Scherlis}
\label{dimer}
\end{figure}

Fig.~\ref{dimer} (upper panel) shows the binding energy
for the [TCNE]$_2^{2-}$ in dichloromethane as a function of the separation
between the [TCNE]$^{-\cdot}$ fragments. At every point, all coordinates were
relaxed while freezing the intradimer distance. The curve presents a steep minimum 
at 3.04 $\AA$, with a barrier to dissociation of nearly 4 kcal/mol.
The grouping of two monomers inside a single cavity, of an area smaller
than the one corresponding to two separate cavities containing one monomer each,
is energetically favored by the surface tension of the
solvent. Thus, if the contribution of the cavitation energy to the solvation is 
not considered, the binding results weaker, as seen in Fig.~\ref{dimer}.
The surface of the cavity, plotted in the lower panel, 
increases gradually as the monomers are pulled
apart, until the solvation cavity splits in two at around 5 $\AA$.
(this is a case in which different $\beta$ in the parametrization could
account for the distinctive ability of solvents to penetrate narrow spaces).
Beyond this point the total surface remains constant as each 
[TCNE]$^{-\cdot}$ unit occupies 
a separate cavity, and the two curves in the top panel merge.
The ground state of the system is a singlet for distances up to
4.0 $\AA$, whereas at larger separations the spins of the fragments
are no longer paired, conforming to a triplet state.

A value of -1.1 kcal/mol is obtained for the binding energy
between the monomers. Such a value is
underestimated with respect to the experimental dimerization enthalpy,
$\Delta H_D$, reported in the range of -6.9 -- -9.8 kcal/mol in
dichloromethane.\cite{kochi} The disagreement can
be partially attributed to the inability of DFT to fully account for
the correlation energy involved in the $\pi$*--$\pi$* bond, and also,
to some extent, to the effect of the counterions present in the
solution, which differentially stabilize [TCNE]$_2^{2-}$ compared to two
[TCNE]$^{-\cdot}$ anions. This effect has been advocated
in a recent study\cite{simons} of the
interaction of two [TCNE]$^{-\cdot}$ fragments in 
tetrahydrofuran ($\epsilon$=7.58) using PCM at the MP2 level, to explain
why the dimer was found metastable by 9.7 kcal/mol with respect
to the isolated monomers---the experimental estimate for
$\Delta H_D$ being -8 kcal/mol in 
2-methyl-tetrahydrofuran.\cite{chang} The binding energy
curve presented in that work exhibited a broad minimum extending from
3.1 to 3.7 $\AA$, a separation range substantially larger than the one 
observed in the solid state.
Our own PCM calculations in dichloromethane, using PBE in combination 
with the 6-311+G(d,p) Gaussian basis set, yield a metastable dimer with an
interaction energy of 3.2 kcal/mol and an equilibrium distance of 3.00 $\AA$.

\begin{figure} \centerline{
\rotatebox{-90}{\resizebox{3in}{!}{\includegraphics{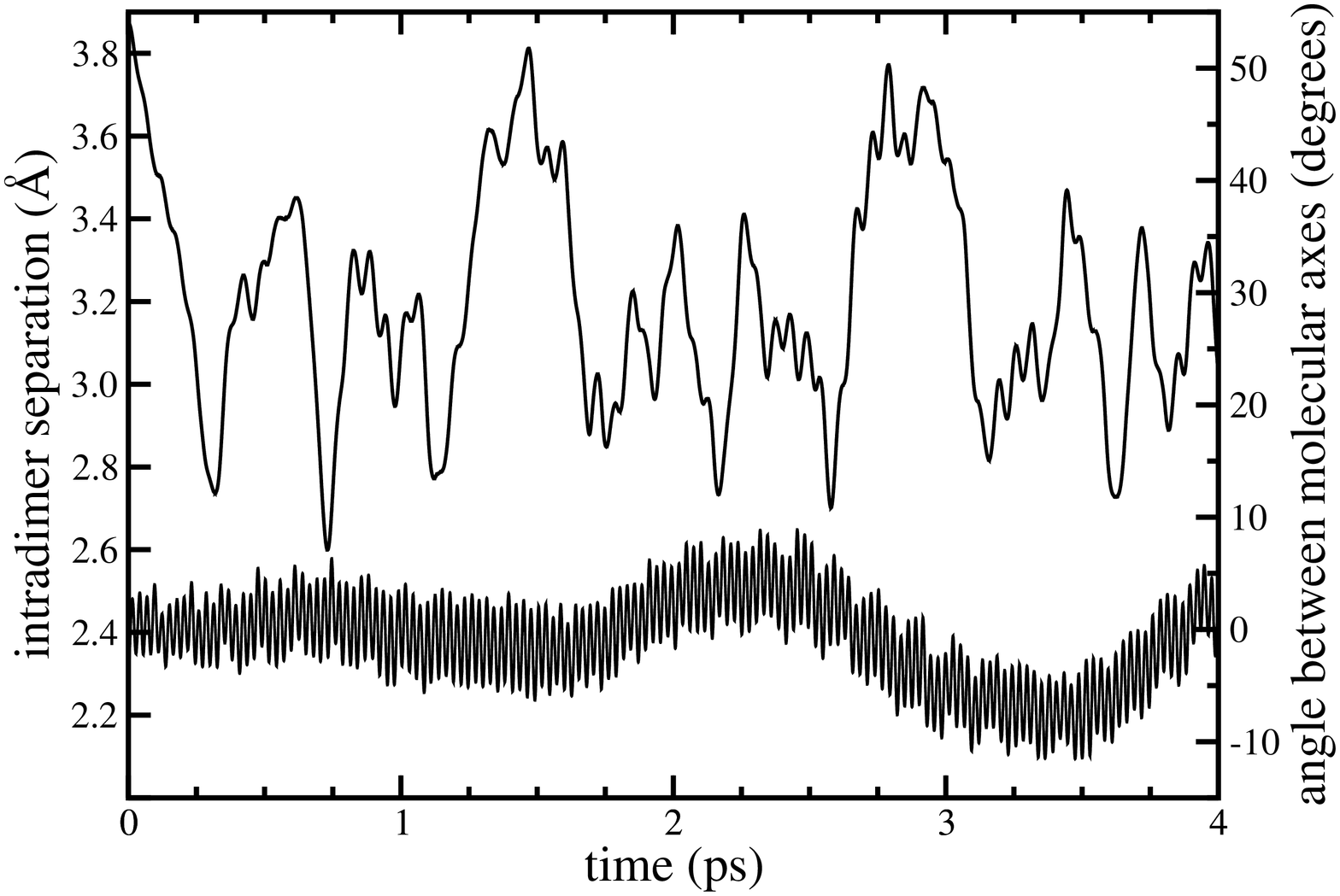}}}
}
\caption{D. Scherlis}
\label{md}
\end{figure}

Temperature dependence investigations in solution
indicate that the dissociated [TCNE]$^{-\cdot}$ anions
are the predominant species at ambient conditions, and that the concentration 
of the dimer rapidly grows as the temperature goes down.\cite{kochi, sesto2}
Car-Parrinello molecular dynamics simulations of the [TCNE]$_2^{2-}$ dimer were 
performed in dichloromethane at 250 K, with the temperature 
controlled by the Nose-Hoover thermostat on the ions.
A time step of 0.288 fs and an electronic mass of 400 a.u. were used.
In Fig.~\ref{md}, we monitor the evolution of two structural parameters
which serve as descriptors of the [TCNE]$^{-\cdot}$-- [TCNE]$^{-\cdot}$
bonding. The intradimer separation, departing from a value of
3.9 $\AA$ corresponding to an initially elongated dimer, drops
to nearly 2.7 $\AA$ and then describes large oscillations in 
the order of 1 $\AA$ around the equilibrium distance.
The second parameter, corresponding to the C=C$\cdots$C=C dihedral 
angle formed by the two [TCNE]$^{-\cdot}$ anions, provides
a measure of the alignment between the monomers: if this
angle is zero the anions lay parallel. Fig.~\ref{md}
shows that this is not the case most of the time.
Rapid oscillations of an average amplitude of 6$^\circ$
take place around the equilibrium angle. During most of the second part
of the run the oscillations are not necessarily centered around zero,
which is indicative of the relatively lax nature of the bond.

\begin{figure} \centerline{
\rotatebox{-90}{\resizebox{3in}{!}{\includegraphics{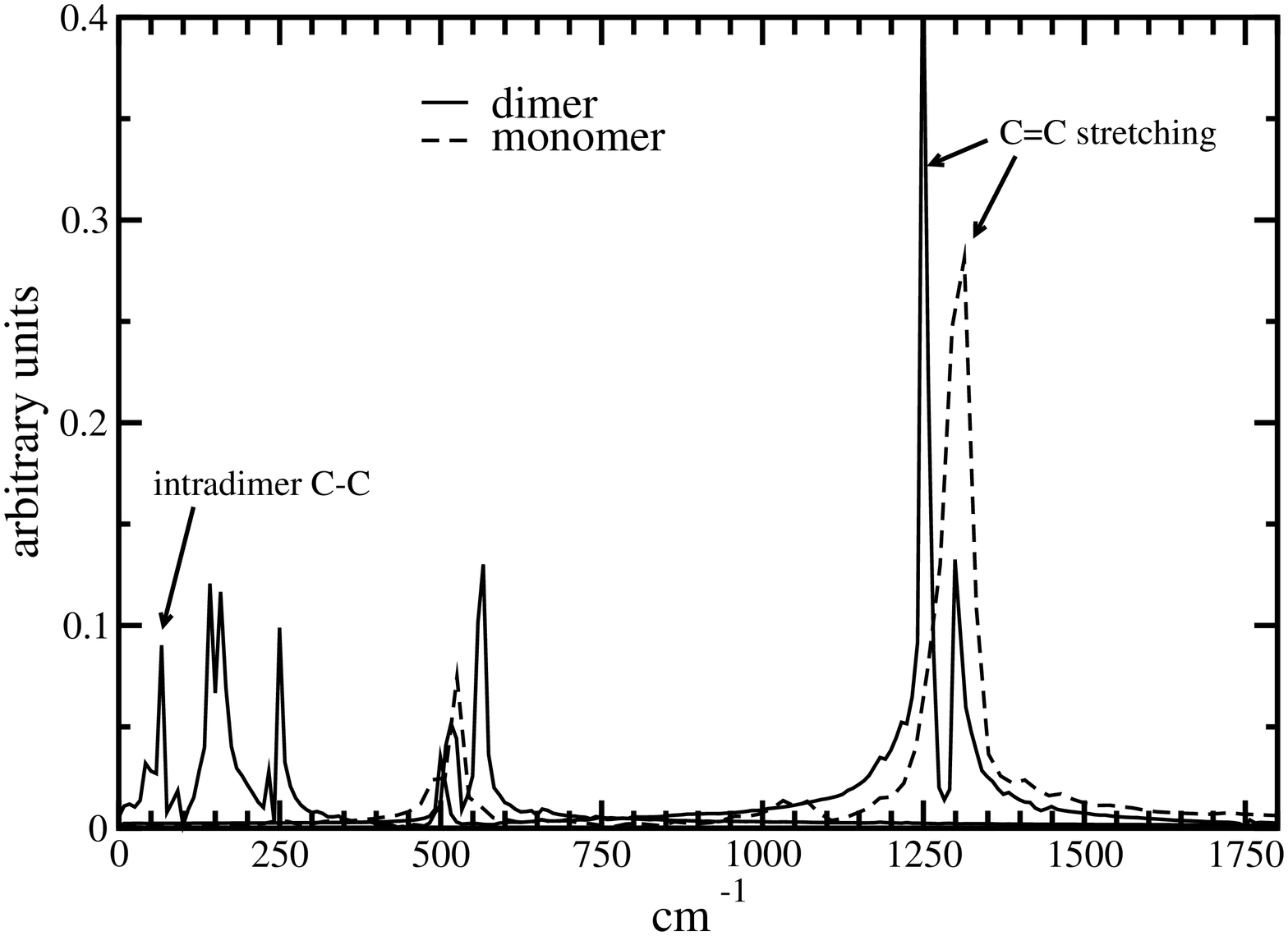}}}
}
\caption{D. Scherlis}
\label{spectrum}
\end{figure}

The length of the simulation is enough to reveal some distinctive features
of the frequency spectrum of the system in the IR region. The continuous
line in Fig.~\ref{spectrum} shows the Fourier transform of the 
velocity-velocity correlation functions corresponding to two pairs of atoms
in the dimer. The first pair consists of the two carbon atoms involved
in the C=C bond. The autocorrelation function of the relative velocity 
between these two centers originates an intense peak corresponding to the
C=C stretching at 1250 cm$^{-1}$. The same mode resolved in the case of
the monomer (dashed line) shows up at 1310 cm$^{-1}$. In the solid state,
experimental C=C stretching frequencies
of 1364 and 1421 cm$^{-1}$ have been reported for the dimer and
the monomer respectively.\cite{sesto2} Such discrepancies between
our results and the experimental numbers are expected, 
given the distinct conditions in the solid and liquid environments,
the difference in the temperatures at which the spectroscopic and
the computational data were collected, the use of DFT, and the slight
downshift in ionic frequencies in Car-Parrinello dynamics.\cite{tangney}
However, we note that the shift of 60 cm$^{-1}$ in going from the 
monomer to the dimer is nicely reproduced by our simulations.

In an attempt to characterize the frequency of the intradimer
$\pi$*--$\pi$* bonding, we have also analyzed the relative-velocity
autocorrelation function for the
two carbon atoms forming the bond, one atom pertaining to each monomer.
The frequency spectrum of this function yields the four groups
of signals appearing below 600 cm$^{-1}$ in Fig.~\ref{spectrum},
the assignment of which is less evident
than in the case of the C=C stretching. Although we are unable to
unambiguously identify all these frequencies, Fourier
transform analysis of the autocorrelation
function for the velocity of the center of
mass of the two fragments (data not shown) points to the lowest
frequency emerging in the spectrum, at 65 cm$^{-1}$, as the
one related to the intradimer vibration. To the best of our knowledge,
no experimental data is available for this mode. Interestingly enough,
though, the aforementioned theoretical study based on PCM and
MP2,\cite{simons} predicted an inter-fragment vibrational frequency
of 60 cm$^{-1}$ by solving the one-dimensional Scr\"{o}dinger
equation on the potential energy surface calculated for the
interaction between the [TCNE]$^{-\cdot}$ anions.

\section{Final remarks}

The electrostatic-cavitation model described in this work enables
Car-Parrinello molecular dynamics simulations in a continuum solvent
for large finite systems, and shows a level of
accuracy as good as that offered by state-of-the-art
quantum chemistry solvation schemes. Additionally,
our model is suited for the treatment of periodic systems
in solution, representing a powerful tool for the study of solid-liquid
interfaces, solvated polymers, and in general extended systems
in contact with a solution. Further improvements will be the subject
of future work, especially the incorporation of the dispersion-repulsion
effects, which become increasingly important with the size of
the solute. The method of references ~\cite{disp1} and~
\cite{disp2} is an attractive choice,
although other possible approaches derived from first-principles
and employing a minimal number of parameters are
also envisioned.

Our cavitation energy, defined in a simple and physical way,
can be straightforwardly implemented in plane-waves or real space codes.
Interestingly, such definition turned out to be
in remarkable agreement with the values
provided by more complex algorithms reliant on large sets of parameters.

The real time study of the pairing
of [TCNE]$^{-\cdot}$ constitutes the first dynamical ab-initio
investigation of dimerization phenomena in solution, of which the
formation of the [TCNE]$_2^{2-}$ is just one example.
The binding of charged radicals in solution is relevant to
a broad field of research in organic and materials chemistry,
and proper consideration of the cavitation contribution turns out to be a
central ingredient for an accurate atomistic description.

\section{Acknowledgments}
The authors thank Patrick Sit for his help in post-processing the velocity
autocorrelation functions, and for useful discussions.
This research was supported by the MURI Grant DAAD 19-03-1-0169,
and by the Institute of Soldier
Nanotechnologies, contract DAAD-19-02-D0002, with the U.S. Army Research Office.

\section{Appendix}

We sumarize here the relevant steps to calculate energies and forces in
the framework of
pseudopotential codes in periodic boundary conditions,
highlighting the additional terms
arising from the electrostatic embedding.
Leaving aside the exchange-correlation energy and the non-local
term of the pseudopotential, the electrostatic problem in a system
of pseudo-ions (nuclei plus core electrons) and 
valence electrons may be written\cite{pasq-galli}
\[
E=\sum_{I<J} \frac{Z_I Z_J}{R_{IJ}} + \sum_I \int \rho_e(r)
v_{loc}(r-R_I)dr\]
\begin{equation}
+ \frac{1}{2} \int \int 
\frac{\rho_e(r) \rho_e(r')}{|r-r'|}dr dr'
\end{equation}
The first term on the right in Eq.~(16) accounts for repulsion between
pseudo-ions, the second is the interaction between these ions and 
the valence electron density, and the third is the Coulombic integral
between valence electrons.
Let $\rho_I(r-R_I)$ be a Gaussian distribution of negative
sign that integrates to the total charge of the pseudo-ion
(note that the electronic charge is defined here as positive).
Adding and subtracting $\sum_I \rho_I(r-R_I)$ from $\rho_e(r)$ in
the third term we obtain

\[ E= \frac{1}{2} \int \int [\rho_e(r)+\sum_I \rho_I(r-R_I)]
[\rho_e(r')\]
\[+ \sum_I \rho_I(r'-R_I)] \frac{1}{|r-r'|}dr dr' \]
\[- \int \int \rho_e(r) [\sum_I \rho_I(r'-R_I)] \frac{1}{|r-r'|}dr dr'\]
\[-\frac{1}{2} \sum_{IJ} \int \int \rho_I(r-R_I) \rho_J(r'-R_J)
\frac{1}{|r-r'|}dr dr'  \]
\begin{equation}
+ \sum_I \int \rho_e(r) v_{loc}(r-R_I)dr +
\sum_{I<J} \frac{Z_I Z_J}{R_{IJ}}
\end{equation}
The first term on the right is the Hartree energy $E_H$ of a pseudopotential
code. Introducing the following definitions:
\[ E_H= \frac{1}{2} \int \int [\rho_e(r)\]
\[+\sum_I \rho_I(r-R_I)]
[\rho_e(r')+ \sum_I \rho_I(r'-R_I)] \frac{1}{|r-r'|}dr dr' \]
\[ E_{ps}=\sum_I \int \rho_e(r) [v_{loc}(r-R_I)+v_I(r-R_I)]dr,\]
\[with~v_I(r)=-\int \frac{\rho_I(r')}{|r-r'|}dr' \]
\[E_{sr}=-\sum_{I<J} \int \int \rho_I(r-R_I) \rho_J(r'-R_J)
\frac{1}{|r-r'|}dr dr'\] \[+ \sum_{I<J} \frac{Z_I Z_J}{R_{IJ}} \]
\[E_{self}=-\frac{1}{2} \sum_{I} \int \int \rho_I(r-R_I) \rho_I(r'-R_I)
\frac{1}{|r-r'|}dr dr' \]
it is possible to write the total energy as:
\begin{equation}
E=E_H + E_{ps} + E_{sr} + E_{self}
\end{equation}
The pseudo-ions density $\rho_I$ is defined as:
\begin{equation}
\rho_I(r-R_I)=-\frac{Z_I}{(R_{I}^c)^3} \pi^{-\frac{3}{2}}
\exp \biggl(-\frac{|r-R_I|^2}{(R_{I}^c)^2}\biggr)
\end{equation}
where $R_{I}^c$ determines the width of the Gaussian associated with
the site $I$. Under such definition $E_{self}$ and $E_{sr}$ can
be evaluated analytically. In particular, $E_{self}$ is a constant
not dependent on the atomic positions:
\begin{equation}
E_{self}=-\frac{1}{\sqrt{2\pi}} \sum_I \frac{Z_I^2}{R_{I}^c}
\end{equation}
\begin{equation}
E_{sr}= \sum_{I<J} \frac{Z_I Z_J}{R_{IJ}} erfc \biggl(
\frac{R_{IJ}}{\sqrt{(R_I^c)^2+(R_J^c)^2}} \biggr)
\end{equation}
The remaining pseudopotential term $E_{ps}$ is computed in reciprocal
space, after constructing the pseudopotential $v_{loc}^I$ carrying
both contributions from the local pseudopotential $v_{loc}$ and
the smeared core charges potential $v_I$.
\begin{equation}
E_{ps}= \sum_I \int \rho_e(r) v_{loc}^I(r) dr
\end{equation}
\[
v_{loc}^I(r)=v_{loc}(r)+v_I(r)=v_{loc}(r)-\int 
\frac{\rho_I(r')}{|r-r'|}dr'\]
\begin{equation}
=v_{loc}(r)-\frac{Z_I}{r} erf \biggl( \frac{r}{R_I^c} \biggr)
\end{equation}

The ionic forces can be obtained from the energies above (plus the
non-local pseudopotential term, which will be omitted for
simplicity). The Hellmann-Feynman theorem---i.e. the stationariety
of the total energy with respect to $\psi$---gives
\begin{equation}
F_I=-\frac{dE}{dR_I}=-\frac{\partial E}{\partial R_I}-\sum_j
\frac{\delta E}{\delta |\psi_j\rangle} \frac{\delta |\psi_j\rangle}
{\partial R_I}=\frac{\partial E}{\partial R_I}
\end{equation}
(real wavefunctions are assumed). Thus,
\begin{equation}
-\frac{dE}{dR_I}=-\frac{\partial E}{\partial R_I}=
-\frac{\partial E_H}{\partial R_I}-
\frac{\partial E_{ps}}{\partial R_I}-\frac{\partial E_{sr}}{\partial R_I}-
\frac{\partial E_{self}}{\partial R_I}
\end{equation}
Note that the partial derivatives of the
individual terms in the Hamiltonian do not correspond to the
total derivatives. For example:
\[ \frac{\partial E_H(R_1,R_2,...,R_n)}{\partial R_n} \neq
\frac{dE}{dR_I}\] \[ = \lim_{\epsilon \to 0}
\frac{E_H(R_1,R_2,...,R_n+\epsilon)-
E_H(R_1,R_2,...,R_n-\epsilon)}{2\epsilon} \]
$E_{self}$ does not depend on $R_I$ and therefore does not contribute to
the forces, whereas the derivative for $E_{sr}$ can be obtained analytically.

The derivative of $E_{ps}$ results
\begin{equation}
\frac{\partial E_{ps}}{\partial R_I}=\sum_I \int \rho_e(r) 
\frac{\partial v_{loc}^I(r)}{\partial R_I} dr
\end{equation}
where the term $\partial v_{loc}^I(r)/\partial R_I$ is
straightforward in the reciprocal space:
\[ v_{loc}^I(r-R_I)= \sum_{\bf G} \tilde v_G e^{i{\bf G r}}
e^{-i{\bf G R_I}} \]
\begin{equation}
\frac{\partial}{\partial R_I}v_{loc}^I(r-R_I)=
\sum_{\bf G} -i {\bf G} \tilde v_G e^{i{\bf G r}} e^{-i{\bf G R_I}}
\end{equation}
with $\tilde v_G$ the coefficients of the Fourier expansion for 
$v_{loc}^I(r)$.

Finally, to obtain the contribution from $E_H$, the Hartree
energy is recast as:
\[ E_H= \frac{1}{2} \int \int \bigg [ \rho_e(r)\rho_e(r')+\sum_I \rho_I(r-R_I)\sum_I \rho_I(r'-R_I) \] \[
+2\rho_e(r)\sum_I \rho_I(r'-R_I) \bigg ] \frac{1}{|r-r'|}dr dr' \]
whose derivative with respect to the atomic positions is:
\[ \frac{1}{2} \int \int \bigg [2\sum_I \rho_I(r-R_I) \biggl (\frac{\partial}{\partial R_I}
\sum_I \rho_I(r'-R_I) \biggr )\] \[+2\rho_e(r) \biggl (\frac{\partial}{\partial R_I}
\sum_I \rho_I(r'-R_I) \biggr) \bigg ] \frac{1}{|r-r'|}dr dr' \]
\[ = \int \int \biggl (\frac{\partial}{\partial R_I} \sum_I \rho_I(r'-R_I) \biggr)\] \[ \times
\biggl (\rho_e(r)+\sum_I \rho_I(r-R_I) \biggr ) \frac{1}{|r-r'|}dr dr' \]
Hence, if $\rho_{tot}(r)=\rho_e(r)+\sum_I \rho_I(r-R_I)$, the contribution from $E_H$
turns out to be
\begin{equation}
\frac{\partial E_H}{\partial R_I}=\int \int \frac{\rho_{tot}(r)}{|r-r'|}
\bigg (\frac{\partial}{\partial R_I} \sum_I \rho_I(r'-R_I) \bigg ) dr dr'
\end{equation}
where the term $\partial \sum_I \rho_I(r'-R_I)/\partial R_I$ is obtained
in Fourier space in the same fashion as in Eq.~(27).
The ratio $\int dr' \rho_{tot}(r)/|r-r'|$ is the Hartree potential $V_H$,
which can be computed in the reciprocal space from the expansion for $\rho_{tot}(r)$.
\[ \rho_{tot}(r)= \sum_{\bf G} \tilde \rho_G e^{i{\bf G r}},~~~~~
V_H=\sum_{\bf G} \tilde \beta_G e^{i{\bf G r}} \]
\[ \nabla^2 V_H=-4\pi \rho_{tot} \Rightarrow \tilde \beta_G = -\frac{4\pi}{{\bf G}^2}
\tilde \rho_G \]
\begin{equation}
V_H= \sum_{\bf G} \frac{-4\pi}{{\bf G}^2} \tilde \rho_G e^{i{\bf G r}}
\end{equation}

In the case of the continuum solvent implementation, $V_H$
is replaced by $\frac{\delta E_{es}}{\delta \rho}({\bf r})$ according
to Eq.~(5) and (6) of the main text. The electrostatic
contribution to the energy originated in the dielectric medium
is computed as
\begin{equation}
E_{es}=\int \rho_e(r) \phi(r) dr
\end{equation}
where $\phi(r)$ is the electrostatic potential obtained using the
multigrid in Eq.~(3). The Hartree term $E_H$ is thus replaced
by $E_{es}$ in the calculation of the total energy.
The cavitation energy
is accounted properly in the total energy by adding
$\gamma S$, which functional derivative (Eq.~(14)) is included
in the Kohn-Sham potential---as the
Hellmann-Feynman theorem applies for that term.

\newpage

\begin{table}
\caption{Solvation and cavitation free energies (kcal/mol) for
neutral solutes in water, calculated 
with this model and with PCM as implemented in Gaussian 03.}

%\vskip .1in
\begin{center}
\begin{ruledtabular}
\begin{tabular}{lccccc}
 & \multicolumn{3}{c} {$\Delta G_{sol}$} & \multicolumn{2}{c} {$\Delta G_{cav}$} \\
\cline{2-4} \cline{5-6}
 & Expt.\cite{sol1}$^-$\cite{sol3} & 
This model & PCM & This model & PCM \\
 \cline{2-4} \cline{5-6}
H$_2$O & -6.3 & -8.4 & -5.4 & 5.7 & 5.7 \\
NH$_3$ &  -4.3 & -3.2 & -1.6 & 6.6 & 6.6 \\
CH$_4$ & 2.0 & 5.4 & 6.9 & 7.5 & 10.0 \\
CH$_3$OH & -5.1 & -3.6 & -0.8 & 9.0 & 9.6 \\
CH$_3$COCH$_3$ & -3.9 & -1.7 & 3.5 & 13.7 & 14.3 \\
HOCH$_2$CH$_2$OH & -9.3 & -9.3 & -6.7 & 13.0 & 12.3 \\
CH$_3$CONH$_2$ & -9.7 & -10.5 & -4.6 & 12.7 & 12.8 \\
CH$_3$CH$_2$CO$_2$H & -6.5 & -6.0 & -2.4 & 14.8 & 14.6 \\
mean unsigned error & & 1.5 & 4.0 & &  \\
max. unsigned error & & 3.4 & 7.4 & & \\
\end{tabular}
\end{ruledtabular}
\end{center}
\end{table}
                                                                                
Table P.1

~~~
\newpage
~~~
\newpage
\begin{table}
\caption{Solvation and cavitation free energies (kcal/mol) for
ionic solutes in water, calculated
with this model and with PCM as implemented in Gaussian 03.}
                                                                                
%\vskip .1in
\begin{center}
\begin{ruledtabular}
\begin{tabular}{lccccc}
 & \multicolumn{3}{c} {$\Delta G_{sol}$} & \multicolumn{2}{c} {$\Delta G_{cav}$} \\
\cline{2-4} \cline{5-6}
 & Expt.\cite{sol1}$^-$\cite{sol3} &
This model & PCM & This model & PCM \\
\cline{2-4} \cline{5-6}
Cl$^-$ & -75 & -66.9 & -72.6 & 7.9 & 5.8 \\
NO$_3^-$ & -65 & -57.8 & -62.6 & 10.5 & 9.7 \\
CN$^-$ & -75 & -64.8 & -70.2 & 8.4 & 7.0 \\
CHCl$_2$CO$_2^-$ & -66 & -74.7 & -53.5 & 16.3 & 15.7 \\
Ag$^+$ & -115 & -110.0 & -102.3 & 5.7 & 4.0 \\
CH$_3$NH$_3^+$ & -73 & -81.0 & -65.1 & 9.4 & 10.2 \\
CH$_3$C(OH)CH$_3^+$ & -64 & -70.6 & -55.2 & 13.5 & 14.4 \\
C$_5$H$_5$NH$^+$ (pyridinium) & -58 & -60.8 & -59.0 & 15.0 & 13.9 \\
mean unsigned error & & 7.1 & 6.6 & &  \\
max. unsigned error & & 9.2 & 12.7 & & \\
\end{tabular}
\end{ruledtabular}
\end{center}
\end{table}
                                                                             
Table P.2
\newpage
~~~
\newpage

\noindent{\bf Figure Captions:}\\
                                                                             
\noindent{\bf Figure 1.}:
Cavity area of a water molecule as a function of
$\Delta$ (thickness parameter used to evaluate the area, see text) for 
several values of the electronic density threshold $\rho_0$.

\noindent{\bf Figure 2.}:
Total energy of the NO$_3^-$ anion as a function of the
inverse of the lattice parameter, computed in vacuum, in solution,
and in vacuum with the Makov-Payne correction up to the leading order.

\noindent{\bf Figure 3.}:
Cluster of D$_2$O molecules used as starting configuration
in the molecular dynamics simulations which results are reported
in Fig. 4.

\noindent{\bf Figure 4.}:
Total and potential energies (top) as a function of time
in a molecular dynamics
simulation of a cyclic tetramer of heavy water in aqueous solution.
The total energy contains the contribution of the
Nose-Hoover thermostat.
The four curves starting at the bottom of the graph represent
the evolution of the intermolecular O$\cdots$H distance between 
the atoms initially involved in hydrogen bonds.

\noindent{\bf Figure 5.}:
Optimized structure of a dimer of [TCNE]$^{-\cdot}$ in
dichloromethane, enclosed by an electronic density
isosurface at 0.00078 e delimiting the solvation cavity. 
Carbon atoms in light gray and nitrogen atoms in dark.

\noindent{\bf Figure 6.}:
Upper panel: binding energy of two [TCNE]$^{-\cdot}$ anions in 
dichloromethane as a function of its separation, calculated
with only the electrostatic contribution to the solvation energy, 
and with both the electrostatic and cavitation contributions.
Lower panel: area of the solvation cavity as a function of the
separation between the [TCNE]$^{-\cdot}$ anions. Above 5 $\AA$
the cavity splits, and the plotted values correspond to the area
of two cavities containing one [TCNE]$^{-\cdot}$ each.

\noindent{\bf Figure 7.}:
Time evolution of the intradimer separation (top)
and the angle determined by the central C=C axes of the two monomers
(bottom) during a molecular dynamics
simulation of [TCNE]$_2^{2-}$ in dichloromethane.

\noindent{\bf Figure 8.}:
Characteristic frequencies of the TCNE monomer and dimer extracted from 
the velocity autocorrelation functions for selected pairs of atoms.

\newpage
                                                                             
\newpage 
                                                                                
\newpage
                                                                                        
\newpage
                                                                                        
\newpage

\newpage

\newpage

\newpage
                                                                                                                             
\begin{thebibliography}{10}
\bibitem{nota1}
The remaining contributions---thermal energy
and $P\Delta V$---are usually negligible and they are rarely considered
in continuum approaches.\cite{crtomasi}
\end{thebibliography}
\end{document}